\documentstyle[preprint,aps,epsf]{revtex}                  
\tightenlines
\begin{document}
\pagestyle{empty}                                      
\preprint{
\font\fortssbx=cmssbx10 scaled \magstep2
\hbox to \hsize{
\hfill$\raise .5cm\vtop{              
                \hbox{NCTU-HEP-9804}}$}
}
\draft
\vfill
\title{Gauge Independent Effective Potential and the Higgs Boson Mass Bound}

\vfill
\author{Guey-Lin Lin and Tzuu-Kang Chyi}

\address{
\rm Institute of Physics, National Chiao-Tung University,
Hsinchu, Taiwan, R.O.C.
}

%
%
\vfill
\maketitle
\begin{abstract}
We introduce the Vilkovisky-DeWitt formalism
for deriving the lower bound of the Higgs boson mass.  
We illustrate the formalism with a simplified 
version of the Standard Electroweak Model, where all charged boson fields
as well as the bottom-quark field are disregarded. The effective potential
obtained in this approach is gauge independent. 
We derive from the effective potential the 
mass bound of the Higgs boson. The result is compared 
to its counterpart obtained from the ordinary
effective potential.   
\end{abstract}
%
%
%
%
\pagestyle{plain}

\section{Introduction}

The effective potentials in quantum field theories are 
off-shell quantities. Therefore, in gauge field theories,  
effective potentials are gauge-dependent as  
pointed out 
by Jackiw in the early seventies\cite{jackiw}. This property  
caused concerns on the physical
significance of effective potentials. In a work by
Dolan and Jackiw \cite{DJ}, the effective potential of scalar
QED was calculated in a set of $R_{\xi}$ gauges. It was concluded that only 
the limiting unitary gauge gives a sensible result on
the symmetry-breaking behaviour of the theory. 
This difficulty was partially resolved 
by the work of Nielsen\cite{niel}.
In his paper, Nielsen derived 
the following identity governing the behaviour of effective potential 
in a gauge 
field theory:
\begin{equation}
(\xi{\partial \over \partial \xi}+C(\phi,\xi)
{\partial\over \partial \phi})V(\phi, \xi)=0,
\end{equation}    
where $\xi$ is the gauge-fixing parameter, $\phi$ is the
order-parameter of the effective potential, and $C(\phi, \xi)$ is the 
Green's function for certain composite operators containing a ghost field.  
The above identity implies that, for different $\xi$, 
the local extrema of $V$ are located along 
the same characteristic
curve on $(\phi,\xi)$ plane, which  satisfies 
$d\xi={d\phi\over C(\phi,\xi)/\xi}$. 
Hence covariant gauges with different $\xi$ are equally
good for computing $V$. On the other hand, a choice of the multi-parameter
gauge $L_{gf}=-{1\over 2 \xi}
(\partial_{\mu}A^{\mu}+\sigma \phi_1 +\rho \phi_2)^2$\cite{DJ},
with $\phi_{1,2}$ the components of the scalar field, would break the 
homogeneity of Eq. (1)\cite{niel}. 
Therefore an effective potential calculated in such a 
gauge does not have a physical significance. 

Recently, it was pointed out\cite{LW} that the Higgs boson mass bound, 
which one derives
from the effective potential, is gauge-dependent. 
The gauge dependence resides in the expression for the 
one-loop effective potential.
Boyanovsky, Loinaz and Willey proposed a resolution\cite{BLW} to the 
problem, which is based upon
the {\it Physical
Effective Potential} constructed as the expectation value of 
the Hamiltonian in physical states\cite{BBHL}. 
They computed the {\it Physical Effective Potential} of
an Abelian Higgs model with an axial coupling of the gauge fields to
the fermions. A gauge-independent lower bound for the  
Higgs boson mass is then determined from the effective potential.  
We note that their approach requires the identification
of first-class constraints of the model and a projection to 
the physical states. Such a procedure is not manifestly
Lorentz covariant. Consequently we expect that it is 
highly non-trivial to apply their approach to the Standard Model(SM).
In our work, we shall introduce the Vilkovisky-DeWitt formalism
\cite{vil,dw2} for constructing a gauge-independent effective potential,
and therefore obtain a gauge-independent lower bound for the Higgs boson 
mass.

In the Vilkovisky-DeWitt formalism, fields are treated as vectors 
in the configuration space,
and the {\it affine connection} of the configuration space is identified to 
facilitate the construction of an invariant effective action. Since
this procedure
is completely Lorentz covariant, the computations for the effective potential 
and the effective action are straightforward. 
We shall perform a calculation with respect to a toy model\cite{LSY} 
which disregards all charged boson fields in the SM.
It is easy to generalize our calculations to the full SM case.   
In fact,  
the applicability of Vilkovisky-DeWitt formalism
to non-Abelian gauge theories has been
extensively demonstrated in the literature\cite{rebhan}.       

The outline of this paper is as follows. In Sec. II, we briefly
review the Vilkovisky-DeWitt formalism using the scalar QED as 
an example. We shall illustrate that the effective action of Vilkovisky
and DeWitt is equivalent to the ordinary effective action
constructed in the Landau-DeWitt gauge\cite{FT}. In Sec. III, we calculate 
the effective potential of the simplified standard model, and the
relevant renormalization constants of the theory using the Landau-DeWitt
gauge. 
The effective potential
is then improved by the renormalization group analysis. 
In Sec. IV, the Higgs boson mass bound
is derived and compared to that given by the ordinary
effective potential in the Landau gauge. We conclude in Sec. V, with 
some technical details
discussed in the Appendix.      

\section{Vilkovisky-DeWitt Effective Action of Scalar QED}

The formulation of Vilkovisky and DeWitt is motivated
by the parametrization dependence of the ordinary effective action,
which can be written generically as\cite{kun}
\begin{eqnarray}
\exp{i\over \hbar}\Gamma[\Phi]&=&\exp{i\over \hbar}(W[j]+\Phi^i{\delta \Gamma
\over \delta \Phi^i})\nonumber \\
&=& \int [D\phi]\exp{i\over \hbar}(S[\phi]-(\phi^i-\Phi^i)\cdot
{\delta \Gamma
\over \delta \Phi^i}),
\label{INTEG}
\end{eqnarray}
where $S[\phi]$ is the classical action, and $\Phi^i$ denote the 
background fields.
The dependence on the parametrization   
arises because the quantum fluctuation $\eta^i\equiv (\phi^i-\Phi^i)$ is not a 
vector
in the field configuration space, hence the product
$\eta^i \cdot {\delta \Gamma
\over \delta \Phi^i}$ is not a scalar under a reparametrization of fields. 
The remedy to this problem is to replace $\eta^i$
with a two-point function $\sigma^i(\Phi,\phi)$ \cite{vil,dw2,com1} 
which, at the
point $\Phi$, 
is tangent to the geodesic connecting $\Phi$ and $\phi$.
The precise form of $\sigma^i(\Phi,\phi)$ depends on the 
connection of the configuration space, $\Gamma^i_{jk}$. It is easy to 
show that\cite{kun} 
\begin{equation}
\sigma^i(\Phi, \phi)=\eta^i-{1\over 2}\Gamma^i_{jk}\eta^j \eta^k
+ O(\eta^3).
\end{equation}
For scalar QED described
by the Lagrangian:
\begin{eqnarray}
L&=&-{1\over 4}F_{\mu\nu}F^{\mu\nu}+(D_{\mu}\phi)^{\dagger}
(D^{\mu}\phi)\nonumber \\
&-& \lambda (\phi^{\dagger}\phi-\mu^2)^2,
\label{SQED}
\end{eqnarray}
with $D_{\mu}=\partial_{\mu}+ie A_{\mu}$ and $\phi={\phi_1+i\phi_2\over 
\sqrt{2}}$, the connection of the configuration space is given by\cite{vil,kun}
\begin{equation}
\Gamma^i_{jk}= {i\brace j k}+T^i_{jk},
\label{gijk}
\end{equation}
where ${i\brace j k}$ is the Christoffel symbol of the field 
configuration space and $T^i_{jk}$ is a quantity induced by generators of
the gauge transformation. The Christoffel symbol ${i\brace j k}$ can be
computed 
from the following metric tensor of scalar QED:
\begin{eqnarray}
G_{\phi_a(x)\phi_b(y)}&=&\delta_{ab}\delta^4(x-y),\nonumber \\ 
G_{A_{\mu}(x)A_{\nu}(y)}&=&-g^{\mu\nu}\delta^4(x-y),\nonumber \\
G_{A_{\mu}(x)\phi_a(y)}&=&0.
\label{metric}
\end{eqnarray}
According to 
Vilkovisky's prescription\cite{vil}, the metric tensor of the 
field configuration space 
is obtained
by differentiating twice with respect to the fields in the 
kinetic Lagrangian. 
For the above metric tensor, we have ${i\brace jk}=0$ since each component
of the tensor is
field-independent. 
However, the
Christoffel symbol would be non-vanishing 
if one parametrizes Eq. (\ref{SQED})
with
polar variables $\rho$ and $\chi$ such that
$\phi_1=\rho \cos\chi$ and $\phi_2=\rho \sin\chi$.  
Finally, to determine $T^i_{jk}$, let us specify generators $g^i_{\alpha}$ of 
the scalar-QED gauge transformations: 
\begin{eqnarray}
g^{\phi_a(x)}_y&=&-\epsilon^{ab}e\phi_b(x)\delta^4(x-y),\nonumber \\
g^{A_{\mu}(x)}_y&=&-\partial_{\mu}\delta^4(x-y),
\label{gener}
\end{eqnarray}
where $\epsilon^{ab}$ is a skew-symmetric tensor with $\epsilon^{12}=1$.
The quantity $T^i_{jk}$ is related to the generators $g^i_{\alpha}$
via\cite{vil}
\begin{equation}
T^i_{jk}=-B^{\alpha}_jD_kg^i_{\alpha}+
{1\over 2}g^l_{\alpha}D_lg^i_{\beta}
B^{\alpha}_jB^{\beta}_k+ j\leftrightarrow k,
\label{tijk}
\end{equation}  
where $B^{\alpha}_k=N^{\alpha\beta}g_{k\beta}$ with
$N^{\alpha\beta}$ being the inverse of $N_{\alpha\beta}\equiv 
g^k_{\alpha}g^l_{\beta}G_{kl}$. The expression for $T^i_{jk}$
can be easily understood 
by realizing that $i, j,\cdots, l$ are function-space indices, while
$\alpha$ and  $\beta$ are space-time indices. Hence, for example,
\begin{equation}
D_{\phi_1(z)}g^{A_{\mu}(x)}_y={\partial g^{A_{\mu}(x)}_y\over \partial 
\phi_1(z)}+ {A_{\mu}(x)\brace j \; \phi_1(z)}g^j_y,  
\end{equation}
where the summation over $j$ also implies an integration over the space-time
variable in the function $j$.

The one-loop effective action of scalar QED can be calculated from Eq.
(\ref{INTEG}) with each quantum fluctuation $\eta^i$ replaced by
$\sigma^i(\Phi, \phi)$. The result is written as\cite{kun}:
\begin{equation}
\Gamma[\Phi]=S[\Phi]-{i\hbar\over 2}\ln\det G+
{i\hbar\over 2}\ln\det \tilde{D}^{-1}
_{ij},
\label{ACTION}
\end{equation}
where $S[\Phi]$ is the classical action with $\Phi$ denoting 
generically the background fields; $\ln\det G$ arises 
from the function-space measure $[D\phi]\equiv \prod_x d\phi(x)
\sqrt{\det G}$; and $\tilde{D}^{-1}_{ij}$ is the modified inverse-propagator:
\begin{equation}
\tilde{D}^{-1}_{ij}={\delta^2 S\over \delta\Phi^i \delta\Phi^j}
-\Gamma^k_{ij}[\Phi]{\delta S\over \delta \Phi^k}.
\label{INVE}
\end{equation}
To study the symmetry-breaking behaviour of the theory, we focus on
the effective potential which is obtained 
from $\Gamma[\Phi]$ by setting each
background field $\Phi^i$ to a constant. 

The Vilkovisky-DeWitt effective potential of scalar QED has been calculated 
in various gauges and different scalar-field parametrizations 
\cite{FT,kun,rt}.     
The results all agree with one another. In this work, we 
calculate the effective potential and other relevant
quantities in the Landau-DeWitt gauge\cite{com2}, which is characterized by the
gauge-fixing term:
$L_{gf}=-{1\over 2\xi}(\partial_{\mu}B^{\mu}-ie\eta^{\dagger}
\Phi+ie\Phi^{\dagger}\eta)^2$,
with $\xi\to 0$. In $L_{gf}$,
$B^{\mu}\equiv A^{\mu}-A^{\mu}_{cl}$, and $\eta \equiv \phi-\Phi$
are quantum fluctuations while $A^{\mu}_{cl}$ and $\Phi$ are background 
fields. For the scalar fields, we further write 
$\Phi={\rho_{cl}+i\chi_{cl}\over 
\sqrt 2}$ and $\eta={\rho+i\chi\over 
\sqrt 2}$. 
The advantage of performing calculations in the Landau-DeWitt gauge
is that $T^i_{jk}$ vanishes\cite{FT} in this case.
In other words, the Vilkovisky-DeWitt formalism coincides with the 
conventional formalism in the Landau-DeWitt gauge. 

For computing the effective potential, we choose
$A^{\mu}_{cl}=\chi_{cl}=0$, i.e., $\Phi={\rho_{cl}\over \sqrt{2}}$. 
In this set of background fields, 
$L_{gf}$ becomes
\begin{equation}
L_{gf}=-{1\over 2\xi}\left(\partial_{\mu}B^{\mu}\partial_{\nu}B^{\nu}
-2e\rho_{cl}\chi\partial_{\mu}B^{\mu}+e^2\rho_{cl}^2\chi^2\right).
\label{GAUGE}
\end{equation}
We note that $B_{\mu}-\chi$ mixing in $L_{gf}$ is 
$\xi$ dependent, and therefore would not cancell out 
the corresponding mixing term in the classical Lagrangian of 
Eq. (\ref{SQED}). This induces mixed-propagators such as
$<0\vert T(A_{\mu}(x)\chi(y)) \vert 0>$ 
or $<0\vert T(\chi(x)A_{\mu}(y)) \vert 0>$. The Faddeev-Popov ghost
Lagrangian in this gauge reads:
\begin{equation}
L_{FP}=\omega^*(-\partial^2-e^2\rho_{cl}^2)\omega.
\label{FADPOP}
\end{equation}
With each part of the Lagrangian determined, we are ready to
compute the effective potential. Since we choose a field-independent 
flat-metric,
the one-loop effective potential is completely determined by 
the modified inverse-propagators $\tilde{D}^{-1}_{ij}$\cite{grassmann}. From 
Eqs. (\ref{SQED}), (\ref{INVE}), (\ref{GAUGE}) 
and (\ref{FADPOP}), we arrive at
\begin{eqnarray}
\tilde{D}^{-1}_{B_{\mu}B_{\nu}}&=&(-k^2+e^2\rho_0^2)g^{\mu\nu}
+(1-{1\over \xi})k^{\mu}k^{\nu},\nonumber \\
\tilde{D}^{-1}_{B_{\mu}\chi}&=&ik^{\mu}e\rho_0(1-{1\over \xi}),
\nonumber \\
\tilde{D}^{-1}_{\chi\chi}&=&(k^2-m_G^2-{1\over \xi}e^2\rho_0^2),
\nonumber \\
\tilde{D}^{-1}_{\rho\rho}&=&(k^2-m_H^2),\nonumber \\
\tilde{D}_{\omega^*\omega}&=&(k^2-e^2\rho_0^2)^{-2},
\label{PROP}
\end{eqnarray} 
where we have set $\rho_{cl}=\rho_0$, which is a space-time independent 
constant, and defined 
$m_G^2= \lambda (\rho_0^2-2\mu^2)$,
$m_H^2=\lambda (3\rho_0^2-2\mu^2)$.         
Using the definition $\Gamma[\rho_0]=-(2\pi)^4\delta^4(0)V_{eff}(\rho_0)$
along with Eqs. (\ref{ACTION}) and
(\ref{PROP}), and taking the limit $\xi\to 0$, we obtain
$V_{eff}(\rho_0)=V_{tree}(\rho_0)+V_{1-loop}(\rho_0)$ with
\begin{equation}
V_{1-loop}(\rho_0)={-i\hbar\over 2}\int {d^nk\over (2\pi)^n}
\ln\left[(k^2-e^2\rho_0^2)^{n-3}(k^2-m_H^2)(k^2-m_+^2)(k^2-m_-^2)\right],
\label{EFFECTIVE}
\end{equation}
where $m_+^2$ and $m_-^2$ are solutions of the quadratic equation
$(k^2)^2-(2e^2\rho_0^2+m_G^2)k^2+e^4\rho_0^4=0$. One notices that the
gauge-boson's degree of freedom in $V_{1-loop}$ 
has been continued to $n-3$ in order to
preserve the relevant Ward identities. For example, this continuation is  
crucial to ensure the Ward identity which relates
the scalar self-energy to the contribution of the tadpole diagram. 
Our expression for $V_{1-loop}(\rho_0)$
agrees with previous results obtained in the unitary gauge\cite{rt}. 
One could also calculate the effective potential 
in the {\it ghost-free} Lorentz 
gauge with $L_{gf}=-{1\over 2\xi}(\partial_{\mu}B^{\mu})^2$. 
The cancellation of the gauge-parameter($\xi$) dependence in the effective 
potential has been demonstrated in 
the case of massless
scalar QED where $\mu^2=0$\cite{FT,kun}. It can be easily extended to the 
massive case, and the resultant effective potential coincides
with Eq. (\ref{EFFECTIVE}). 
In the Appendix, we will also 
demonstrate the cancellation of gauge-parameter dependence in the 
calculation of Higgs-boson self-energy. The
obtained self-energy will be shown to coincide with its counterpart obtained
from the Landau-DeWitt gauge. We do this not only  
to show that the Vilkovisky-DeWitt formulation coincides with the ordinary 
formulation in the Landau-DeWitt
gauge, but also to illustrate how it gives rise to identical effective action 
in spite of beginning with different gauges.      
   
It is instructive to rewrite Eq. (\ref{EFFECTIVE}) as
\begin{equation}
V_{1-loop}[\rho_0]={\hbar \over 2}\int {d^{n-1}\vec{k}\over (2\pi)^{n-1}} 
\left((n-3)\omega_B(\vec{k})+\omega_H(\vec{k})+\omega_+(\vec{k})
+\omega_-(\vec{k})\right),
\end{equation}
where $\omega_B(\vec{k})=\sqrt{\vec{k}^2+e^2\rho_0^2}$, 
$\omega_H(\vec{k})=\sqrt{\vec{k}^2+m_H^2}$ and $\omega_{\pm}(\vec{k})
=\sqrt{\vec{k}^2+m_{\pm}^2}$. One can see that $V_{1-loop}$ is 
a sum of the zero-point energies of four excitations with masses
$m_B\equiv e\rho_0$, $m_H$, $m_+$ and $m_-$. Since there 
are precisely four physical degrees of freedom in the scalar QED,
we see that the Vilkovisky-DeWitt effective potential does exhibit a 
correct number of physical degrees of freedom. Such a nice property is
not shared by the ordinary effective potential calculated in the
{\it ghost free} Lorentz gauge just mentioned. As will be shown later,
the ordinary effective potential in this gauge  
contains unphysical degrees of freedom.

\section{Vilkovisky-DeWitt Effective Potential of the Simplified Standard
Model}

In this section, we compute the effective potential of the simplified 
standard model where charged boson fields and all fermion fields except 
the top-quark field are disregarded. The gauge interactions for the top quark
and the neutral scalar bosons are
prescribed by the following covariant derivatives\cite{LSY}:
\begin{eqnarray}
D_{\mu}t_L&=&(\partial_{\mu}+ig_LZ_{\mu}-{2\over 3}ieA_{\mu})t_L,
\nonumber \\
D_{\mu}t_R&=&(\partial_{\mu}+ig_RZ_{\mu}-{2\over 3}ieA_{\mu})t_R,
\nonumber \\
D_{\mu}\phi&=&(\partial_{\mu}+i(g_L-g_R)Z_{\mu})\phi,
\end{eqnarray}
where $Z_{\mu}$ and $A_{\mu}$ denote the $Z$ boson and the photon 
respectively; the coupling constants $g_L$ and $g_R$ are given by   
$g_L=(-g_1/2+g_2/3)$ and $g_R=g_2/3$ with $g_1=g/\cos\theta_W$ and
$g_2=2e\tan\theta_W$ respectively. 
The self-interactions of scalar fields are described by the same potential
term as that in Eq. (\ref{SQED}). 
Clearly this toy model exhibits a $U(1)_A\times U(1)_Z$
symmetry where each $U(1)$ symmetry is 
associated with a neutral gauge boson. The $U(1)_Z$-charges of $t_L$, $t_R$
and $\phi$ are related in such a way that the following 
Yukawa interactions are
invariant under $U(1)_A\times U(1)_Z$:
\begin{equation}
L_Y=-y\bar{t}_L\phi t_R-y\bar{t}_R\phi^* t_L.
\end{equation}
Since Vilkosvisky-DeWitt effective action coincides with the 
ordinary effective 
action in the Landau-DeWitt gauge, we thus
calculate the effective potential in this gauge, which is defined by
the following gauge-fixing terms 
\cite{fermion}:
\begin{eqnarray}
L_{gf}=&-&{1\over 2\alpha}(\partial_{\mu}{\tilde Z}^{\mu}+
{ig_1\over 2}\eta^{\dagger}
\Phi-{ig_1\over 2}\Phi^{\dagger}\eta)^2\nonumber \\
&-&{1\over 2\beta}(\partial_{\mu}{\tilde A}^{\mu})^2,
\label{FIXG}
\end{eqnarray}
with $\alpha, \ \beta\to 0$. 
We note that ${\tilde A}^{\mu}$,
${\tilde Z}^{\mu}$ and $\eta$ are quantum fluctuations associated 
with the photon, the $Z$ boson and the scalar boson respectively, i.e.,
$A^{\mu}=A_{cl}^{\mu}+{\tilde A}^{\mu}$,
$Z^{\mu}=Z_{cl}^{\mu}+{\tilde Z}^{\mu}$, and $\phi=\Phi+\eta$ with 
$A_{cl}^{\mu}$, $Z_{cl}^{\mu}$ and $\Phi$ being the background fields.
For computing the effective potential, we take $\Phi$ as 
a space-time-independent constant denoted as $\rho_o$, and set
$A_{cl}^{\mu}=Z_{cl}^{\mu}=0$.  
Following the method 
outlined in the previous section, we obtain the one-loop effective 
potential
\begin{equation}
V_{VD}(\rho_0)={\hbar \over 2}\int {d^{n-1}\vec{k}\over (2\pi)^{n-1}} 
\left((n-3)\omega_Z(\vec{k})+\omega_H(\vec{k})+\omega_+(\vec{k})
+\omega_-(\vec{k})-4\omega_F(\vec{k})\right),
\label{POTENTIAL2}
\end{equation}
where $\omega_i(\vec{k})=\sqrt{\vec{k}^2+m_i^2}$ with
$m_Z^2={g_1^2\over 4}\rho_0^2$, $m_{\pm}^2=m_Z^2+{1\over 2}(m_G^2\pm 
m_G\sqrt{m_G^2+4m_Z^2})$ and $m_F^2\equiv m_t^2={y^2\rho_0^2\over 2}$.
The Goldstone boson mass $m_G$ is defined as before, i.e, $m_G^2=\lambda
(\rho_o^2-2\mu^2)$ with $\mu$ being the mass parameter of the Lagrangian.
One may notice the absence of 
photon contributions in the above effective potential. This is
not surprising since photons do not couple directly to the Higgs boson.

Performing the integration in Eq. (\ref{POTENTIAL2})
and subtracting the infinities with $\overline{MS}$ prescription, we obtain
\begin{eqnarray}
V_{VD}(\rho_0)&=&{\hbar\over 64\pi^2}\left(m_H^4\ln{m_H^2\over \kappa^2} 
+m_Z^4\ln{m_Z^2\over \kappa^2}+m_+^4\ln{m_+^2\over \kappa^2}+
m_-^4\ln{m_-^2\over \kappa^2}-4m_t^4\ln{m_t^2\over \kappa^2}\right)\nonumber \\
&-&{\hbar\over 128\pi^2}\left(3m_H^4+5m_Z^4+3m_G^4+12m_G^2m_Z^2-12m_t^4\right),
\end{eqnarray}
where $\kappa$ is the mass scale introduced in the dimensional regularization.
Although $V_{VD}(\rho_0)$ is obtained in the Landau-DeWitt gauge, we
should stress that any other gauge with non-vanishing $T^i_{jk}$ should lead to
the same result. For later comparisons, let us write down the ordinary 
one-loop effective
potential in the Lorentz gauge(removing the scalar part of Eq. (\ref{FIXG})) 
as follows\cite{DJ}:
\begin{equation}
V_{L}(\rho_0)={\hbar \over 2}\int {d^{n-1}\vec{k}\over (2\pi)^{n-1}} 
\left((n-1)\omega_Z(\vec{k})+\omega_H(\vec{k})+\omega_a(\vec{k},\alpha)
+\omega_b(\vec{k},\alpha)-4\omega_F(\vec{k})\right),
\end{equation}
where 
$\alpha$ is the gauge-fixing parameter and
$\omega_{a,b}(\vec{k},\alpha)=\sqrt{\vec{k}^2+m_{a,b}^2(\alpha)}$ with 
$m_a^2(\alpha)=1/2\cdot (m_G^2+\sqrt{m_G^4-4\alpha m_Z^2m_G^2})$
and $m_b^2(\alpha)=1/2\cdot (m_G^2-\sqrt{m_G^4-4\alpha m_Z^2m_G^2})$.
It is easily seen that there are 6 bosonic degrees of freedom in $V_{L}$,
i.e., two extra degrees of freedom emerge as a result of 
choosing the Lorentz gauge. 
In the Landau gauge, which is a special case of 
the Lorentz gauge with $\alpha=0$, 
there is still one extra
degree of freedom in the effective potential\cite{BBHL}.  
Since 
the Landau gauge is adopted most frequently for computing 
the ordinary effective potential, 
we shall take $\alpha=0$ in $V_L$ hereafter.
Performing the integrations in $V_{L}$ and subtracting the 
infinities, we obtain
\begin{eqnarray}
V_{L}(\rho_0)&=&{\hbar\over 64\pi^2}\left(m_H^4\ln{m_H^2\over \kappa^2} 
+3m_Z^4\ln{m_Z^2\over \kappa^2}+m_G^4\ln{m_G^2\over \kappa^2}
-4m_t^4\ln{m_t^2\over \kappa^2}\right)\nonumber \\
&-&{\hbar\over 128\pi^2}\left(3m_H^4+5m_Z^4+3m_G^4-12m_t^4\right).
\end{eqnarray}
We remark that $V_{L}$ differs from $V_{VD}$ except
at the point of extremum where $\rho_0^2=2\mu^2$. 
At this
point, one has 
$m_G^2=0$ and $m_{\pm}^2
=m_Z^2$, which lead to $V_{VD}(\rho_0=2\mu^2)=V_L(\rho_0^2=2\mu^2)$. 
That 
$V_{VD}=V_L$ at the point of extremum is a consequence of the Nielsen
identity\cite{niel} mentioned earlier. 

To derive the Higgs boson mass bound from $V_{VD}(\rho_0)$ or $V_L(\rho_0)$,
one encounters a breakdown of the perturbation theory at the point
of sufficiently large $\rho_0$ such that, for instance, 
${\lambda\over 16\pi^2}\ln{\lambda\rho_0^2\over \kappa^2}>1$.
To extend the validity of the effective potential to the
large-$\rho_0$ region, the effective potential has to be improved by  
the renomalization group(RG) analysis. Let us denote the effective potential 
as $V_{eff}$ which includes the tree-level contribution 
and quantum corrections. 
The renormalization-scale independence of $V_{eff}$
implies the following equation\cite{cw,BLW}:
\begin{equation}
\left(
-\mu(\gamma_{\mu}+1){\partial \over \partial \mu}
+\beta_{\hat{g}}{\partial
\over \partial \hat{g}}
-(\gamma_{\rho}+1)t{\partial \over \partial t}
+4\right)V_{eff}
(t\rho_0^i,\mu,\hat{g},\kappa)=0.
\end{equation}
where $\mu$ is the mass parameter of the Lagrangian as shown in 
Eq. (\ref{SQED}), and
\begin{eqnarray}
& & \beta_{\hat{g}}=\kappa {d\hat{g}\over d\kappa},\nonumber \\
& & \gamma_{\rho}=-\kappa {d\ln \rho \over d\kappa},\nonumber \\
& & \gamma_{\mu}=-\kappa {d\ln \mu\over d\kappa}, 
\end{eqnarray}
with $\hat{g}$ denoting collectively the coupling constants $\lambda$, $g_1$,
$g_2$ and $y$; $\rho_0^i$ is an arbitrarily chosen initial value for $\rho_0$.
Solving this differential equation gives 
\begin{equation}
V_{eff}(t\rho_0^i,\mu_i,\hat{g}_i,\kappa)=\exp\left(\int_0^{\ln t}
{4\over 1+\gamma_{\rho}(x)}dx\right)V_{eff}(\rho_0^i,\mu(t,\mu_i),
\hat{g}(t,\hat{g}_i),\kappa),
\label{IMPROVE}
\end{equation}  
with $x=\ln(\rho_0^{\prime}/\rho_0^i)$ for an intermediate scale 
$\rho_0^{\prime}$, and
\begin{equation}
t{d\hat{g}\over dt}={\beta_{\hat{g}}(\hat{g}(t))\over 1+\gamma_{\rho}
(\hat{g}(t))} \ {\rm with} \ \hat{g}(0)=\hat{g}_i,
\label{BEGA}
\end{equation} 
\begin{equation}
\mu(t,\mu_i)=\mu_i\exp\left(-\int_0^{\ln t}
{1+\gamma_{\mu}(x)\over 1+\gamma_{\rho}(x)}dx\right).
\label{SCALE}
\end{equation}
To fully determine $V_{eff}$ at a large $\rho_0$, we need to calculate 
the $\beta$ functions of $\lambda$, $g_1$, $g_2$ and $y$, and the anomalous
dimensions $\gamma_{\mu}$ and $\gamma_{\rho}$. It has been demonstrated 
that the $n$-loop effective potential is 
improved by the $(n+1)$-loop $\beta$ and $\gamma$ functions\cite{ka,bkmn}. 
Since  
the effectve potential is calculated to the one-loop order, 
a consistent RG analysis
requires the knowledge of $\beta$ and $\gamma$ functions up to a two-loop
accuracy.
However, as the computations of two-loop $\beta$ and $\gamma$ functions 
are quite involved, we will only improve the tree-level 
effective potential with one-loop $\beta$ and $\gamma$ functions. After all,
the main focus of this paper is to show how to 
obtain a gauge-independent Higgs boson  mass 
bound rather than a detailed calculation of this quantity.    

To compute one-loop $\beta$ and $\gamma$ functions,
we first calculate the renormalization constants $Z_{\lambda}$, $Z_{g_1}$,
$Z_{g_2}$, $Z_y$, $Z_{\mu^2}$ and $Z_{\rho}$, which are defined by
\begin{eqnarray}
\lambda^{bare}&=&Z_{\lambda}\lambda, \;\; g_1^{bare}=Z_{g_1}g_1, \;\;
g_2^{bare}=Z_{g_2}g_2, \nonumber \\
y^{bare}&=&Z_yy, \;\; (\mu^2)^{bare}=Z_{\mu^2}\mu^2, \;\; \rho^{bare}
=\sqrt{Z_{\rho}}\rho.
\end{eqnarray}
In the ordinary formalism of the effective action, 
all of the above renormalization
constants except $Z_{\rho}$ are in fact gauge-independent at 
the one-loop order
in the $\overline{MS}$ scheme. For $Z_{\rho}$,
the result given by the commonly adopted Landau gauge differs 
from that obtained from
the Landau-DeWitt gauge. In Appendix A, we shall reproduce $Z_{\rho}$
obtained in the Landau-DeWitt gauge 
with the general Vilkovisky-DeWitt formulation. 
The calculation of various renormalization constants are straightforward.
In the $\overline{MS}$ scheme, we have(we will set 
$\hbar=1$ from this point on):
\begin{eqnarray}
Z_{\lambda}&=&1-{1\over 128\pi^2\epsilon'}\left({3g_1^4\over \lambda}
-24g_1^2-{16y^4\over \lambda}+32y^2+160\lambda\right),\nonumber \\
Z_{g_1}&=&Z_{g_2}=1-{1\over 216\pi^2\epsilon'}\left({27g_1^2\over 8}
+2g_2^2-3g_1g_2\right), \nonumber \\
Z_y&=&1+{1\over 192\pi^2\epsilon'}\left(9g_1^2+4g_1g_2-24y^2\right),\nonumber
\\
Z_{\mu^2}&=&1+{1\over 128\pi^2\epsilon'}\left({3g_1^4\over \lambda}
-12g_1^2-{16y^4\over \lambda}+16y^2+96\lambda 
\right),\nonumber \\
Z_{\rho}&=&=1+{1\over 32\pi^2\epsilon'}\left(-5g_1^2+4y^2\right),
\label{renc}
\end{eqnarray}
where $1/\epsilon'\equiv 1/\epsilon+{1\over 2}\gamma_E-{1\over 2}\ln(4\pi)$
with $\epsilon=n-4$. The one-loop $\beta$ and $\gamma$ functions resulting
from the above renormalization constants are:
\begin{eqnarray}
\beta_{\lambda}&=&{1\over 16\pi^2}\left({3\over 8}g_1^4-3\lambda g_1^2
-2y^4+4\lambda y^2+20\lambda^2\right),\nonumber \\
\beta_{g_1}&=&{g_1\over 4\pi^2}\left({g_1^2\over 16}-{g_1g_2\over 18}
+{g_2^2\over 27}\right),\nonumber \\
\beta_{g_2}&=&{g_2\over 4\pi^2}\left({g_1^2\over 16}-{g_1g_2\over 18}
+{g_2^2\over 27}\right),\nonumber \\
\beta_y&=&{y\over 8\pi^2}\left(y^2-{3g_1^2\over 8}+{g_1g_2\over 12}\right),
\nonumber \\  
\gamma_{\mu}&=&{1\over 2\pi^2}\left({3\lambda\over 4}+{3g_1^4\over 128\lambda}
-{3g_1^2\over 32}-{y^4\over 8\lambda}+{y^2\over 8}\right),\nonumber \\
\gamma_{\rho}&=&{1\over 64\pi^2}\left(-5g_1^2+4y^2\right).
\label{BETA}
\end{eqnarray}
Similar to what was mentioned earlier, 
all of the above quantities   
are gauge-independent in the $\overline{MS}$
scheme except $\gamma_{\rho}$, the anomalous dimension of the scalar field.
In the Landau gauge of the ordinary formulation,
we have 
\begin{equation}
\gamma_{\rho}={1\over 64\pi^2}\left(-3g_1^2+4y^2\right).
\end{equation}

\section{The Higgs Boson Mass Bound}
The lower bound of the Higgs boson mass can be derived from the 
vacuum instability
condition of the electroweak effective potential\cite{SHER}. 
In this derivation, there exists different criteria 
for determining the instability scale of the electroweak vacuum.
The first criterion is to identify the instability
scale as the critical value of
the Higgs-field strength beyond which the RG-improved tree-level effective 
potential becomes negative\cite{FJSE,SHER2,AI}. To implement this criterion, 
the
tree-level effective potential is improved by the leading\cite{AI} or 
next-to-leading order 
\cite{FJSE,SHER2} renormalization group equations, where one-loop or 
two-loop $\beta$ and $\gamma$ functions are employed.
Furthermore, one-loop corrections to parameters of the effective 
potential are also taken into account\cite{SHER2,AI}. However, the effect of
one-loop effective potential is not considered. 

To improve the above treatment, Casas et. al\cite{ceq}  
considered the effect of RG-improved one-loop effective potential. 
The vacuum-instability scale is then identified as the value of 
the Higgs-field strength
at which the sum of tree-level and one-loop effective potentials vanishes.   
In our subsequent analysis, we will follow this 
criterion except that the one-loop effective potential is not RG-improved.
           
To derive the Higgs boson mass bound, one begins with Eq. (\ref{IMPROVE})
which implies
\begin{equation}
V_{tree}(t\rho_0^i,\mu_i,\lambda_i)={1\over 4}
\chi(t)\lambda(t,\lambda_i)
\left((\rho_0^i)^2-
2\mu^2(t,\mu_i)\right)^2,
\end{equation}
with $\chi(t)=
\exp\left(\int_0^{\ln t}
{4\over 1+\gamma_{\rho}(x)}dx\right)$. Since Eq. (\ref{SCALE}) implies that
$\mu(t,\mu_i)$ decreases
as $t$ increases, we then have $V_{tree}(t\rho_0^i,\mu_i,\lambda_i)
\approx {1\over 4}\chi(t)\lambda(t,\lambda_i)(\rho_0^i)^4$ for a sufficiently 
large $t$. Similarly, the one-loop effective potential 
$V_{1-loop}(t\rho_0^i,\mu_i,\hat{g}_i,\kappa)$ is also proportional to 
$V_{1-loop}(\rho_0^i,\mu(t,\mu_i),\hat{g}(t,\hat{g}_i),\kappa)$ with
the same proportional constant $\chi(t)$. Because 
we shall ignore all running effects in $V_{1-loop}$, we can take 
$\hat{g}(t,\hat{g}_i)=\hat{g}_i$ and $\mu(t,\mu_i)={1\over t}\mu_i$
in $V_{1-loop}$. For
a sufficiently large $t$, $V_{1-loop}$ can also be approximated by its quartic
terms. In the Landau-DeWitt gauge with the choice $\kappa=\rho_0^i$,
we obtain 
\begin{eqnarray}
V_{VD}&\approx&{(\rho_0^i)^4\over 64\pi^2}\left[9\lambda_i^2\ln(3\lambda_i)
+{g_{1i}^4\over 16}\ln({g_{1i}^2\over 4})-y_i^4\ln({y_i^2\over 2})\right.
\nonumber \\
&+&\left. A_+^2(g_{1i},\lambda_i)\ln A_+(g_{1i},\lambda_i)
+A_-^2(g_{1i},\lambda_i)\ln A_-(g_{1i},\lambda_i)\right. \nonumber \\
&-&\left. {3\over 2}(10\lambda_i^2+
\lambda_i g_{1i}^2+{5\over 48}g_{1i}^4-y_i^4)\right],
\label{VVD}
\end{eqnarray}  
where $A_{\pm}(g_1,\lambda)=g_1^2/4+\lambda/2\cdot
(1\pm \sqrt{1+g_1^2/\lambda})$.
Similarly, the effective potential in the Landau gauge is given by
\begin{eqnarray}
V_{L}&\approx&{(\rho_0^i)^4\over 64\pi^2}\left[9\lambda_i^2\ln(3\lambda_i)
+{3g_{1i}^4\over 16}\ln({g_{1i}^2\over 4})
-y_i^4\ln({y_i^2\over 2})\right. \nonumber \\
&+&\left.\lambda_i^2\ln(\lambda_i)
-{3\over 2}(10\lambda_i^2+\lambda_i g_{1i}^2+{5\over 48}g_{1i}^4-y_i^4)\right],
\label{VL}
\end{eqnarray}  
Combining the tree-level and the one-loop effective potentials, we arrive at
\begin{equation}
V_{eff}(t\rho_0^i,\mu_i,\hat{g}_i,\kappa)\approx{1\over 4}\chi(t)
\left(\lambda(t,\lambda_i)+\Delta \lambda(\hat{g}_i)\right)(\rho_0^i)^4,
\end{equation}
where $\Delta \lambda$ represents the one-loop corrections obtained from Eqs.
(\ref{VVD}) or (\ref{VL}). Let $t_{VI}=\rho_{VI}/\rho_0^i$, 
the condition for the vacuum
instability of the effective potential is then\cite{ceq} 
\begin{equation}
\lambda(t_{VI},\lambda_i)+\Delta \lambda(\hat{g}_i)=0.
\label{VI}
\end{equation}  
We note that the couplings $\hat{g}_i$ in $\Delta \lambda$ are evaluated 
at $\kappa=\rho_0^i$, which can be taken as the electroweak scale. Hence
we have 
$g_{1i}\equiv g/\cos\theta_W=0.67$, $g_{2i}\equiv 2e\tan\theta_W=0.31$,
and $y_i=1$. The running coupling $\lambda(t_{VI},\lambda_i)$ also depends upon
$g_1$, $g_2$ and $y$ through $\beta_{\lambda}$, and $\gamma_{\rho}$ 
shown in Eq. (\ref{BETA}).
To 
solve Eq. (\ref{VI}), we first determine the running behaviours of the 
coupling 
constants $g_1$,
$g_2$ and $y$. For $g_1$ and $g_2$, we have
\begin{equation}
t {d\left(g_l^2(t)\right)\over dt}=2g_l(t){\beta_{g_l}
(\hat{g}(t))\over 1+\gamma_{\rho}(\hat{g}(t))}\approx \beta_{g_l^2},
\end{equation}
where $l=1, \ 2$, and 
the contribution of $\gamma_{\rho}$ is neglected in accordance 
with our leading-logarithmic approximation. Also $\beta_{g_l^2}
=g_l^2/2\pi^2\cdot (g_1^2/16-g_1g_2/18+g_2^2/27)$. Although the differential
equations for $g_1^2$ and $g_2^2$ are coupled, they can be easily 
disentangled by observing that $g_1^2/g_2^2$ is a RG-invariant. 
Numerically, we have
$\beta_{g_l^2}=c_lg_l^4$ with $c_1=2.3\times 10^{-3}$ and $c_2=
1.1\times 10^{-2}$. Solving the differential equations gives
\begin{equation}
g_l^{-2}(t)=g_l^{-2}(0)-c_l\ln t.
\end{equation}     
With $g_1(t)$ and $g_2(t)$ determined, the running behaviour of 
$y$ can be calculated analytically\cite{LW}. Given $\beta_{y^2}\equiv 
2y\beta_y=c_3y^4-c_4g_1^2y^2$ with $c_3=2.5\times 10^{-2}$ and
$c_4=8.5\times 10^{-3}$, we obtain
\begin{equation}
y^2(t)=\left[\left({g_1^2(t)\over g_{1i}^2}\right)^{c_4\over c_1}
\left(y_i^{-2}-{c_3\over c_1+c_4}g_{1i}^{-2}\right)+
{c_3\over c_1+c_4}g_1^{-2}(t)
\right]^{-1}.
\end{equation}       

Now the strategy for solving Eq. (\ref{VI}) is to make an initial guess 
on $\lambda_i$, which enters into $\lambda(t)$ and $\Delta \lambda$, and
repeatedly adjust $\lambda_i$ until $\lambda(t)$ completely 
cancells $\Delta \lambda$. 
For $t_{VI}=10^2$(or $\rho_0\approx 10^4$ GeV) which is 
the new-physics scale reachable by
LHC, we find $\lambda_i=4.83\times 10^{-2}$ for the Landau-DeWitt gauge,
and $\lambda_i=4.8\times 10^{-2}$ for the Landau gauge. For a higher 
instability scale such as the scale of grand unification, 
we have $t_{VI}=10^{13}$
or $\rho_0\approx 10^{15}$ GeV. In this case, we find $\lambda_i=3.13\times
10^{-1}$ for both the Landau-DeWitt and Landau gauges. 
The numerical similarity 
between the $\lambda_i$ of each gauge can be attributed to
an identical $\beta$ function for the running of $\lambda(t)$, and
a small difference between the $\Delta \lambda$ of each gauge.
We recall from Eq. (\ref{BEGA}) that the evolutions of $\lambda$ in the above
two
gauges will be different if the effects of next-to-leading logarithms are
taken into account. In that case, the difference between the $\gamma_{\rho}$ 
of each gauge gives rise to different evolutions for $\lambda$. 
For a large $t_{VI}$, one may expect
to see a non-negligible difference between the $\lambda_i$ of each gauge.

The critical value $\lambda_i=4.83\times 10^{-2}$ corresponds to a lower 
bound for the $\overline{MS}$ mass of the Higgs boson. Since $m_H=2\sqrt{
\lambda}\mu$, we have $(m_H)_{\overline{MS}}\geq 77$ GeV. For 
$\lambda_i=3.13\times 10^{-1}$, we have $(m_H)_{\overline{MS}}\geq 196$ GeV.
To obtain the lower bound for the physical mass of the Higgs boson,
finite radiative corrections must be added to the above bounds\cite{LW}. 
We will not pursue these finite corrections any further since we are
simply dealing with a toy model. However we would like to point out 
that such corrections are gauge-independent as ensured by  
the Nielsen identity\cite{niel}.    

\section{Conclusion}
We have computed the one-loop effective potential of 
an Abelian $U(1)\times U(1)$
model in the Landau-DeWitt gauge, which reproduces the result
given by the gauge-independent 
Vilkovisky-DeWitt formulation. One-loop $\beta$ and $\gamma$ functions 
were also computed to facilitate the RG-improvement of the effective 
potential. A gauge-independent lower bound for the Higgs-boson self-coupling
or equivalently the $\overline{MS}$ mass of the Higgs boson was derived.
We compared this bound to that obtained using the ordinary 
Landau-gauge effective potential.
The numerical values of both bounds are
almost identical due to the leading-logarithmic approximation we have taken.
A complete next-to-leading-order analysis 
should better distinguish the two bounds. This improvement as well as
extending the current analysis to the full Standard Model
will be reported in future
publications.  

Finally we would like to comment on the issue of comparing our result 
with that of
Ref.\cite{BLW}. 
So far, we have not found a practical way of 
relating the effective potentials 
calculated in both approaches. In Ref.\cite{BLW}, to achieve 
a {\it gauge-invariant}
formulation, the theory is written in terms of a new set of fields 
which are related to the original fields through non-local transformations.
Taking scalar QED as an example, 
the new scalar field $\phi^{\prime}(\vec{x})$ is related to the 
original field through\cite{BBHL} 
\begin{equation}
\phi^{\prime}(\vec{x})=\phi(\vec{x})\exp\left(ie\int d^3y
\vec{A}(\vec{y})\cdot \vec{\nabla}_yG(\vec{y}-\vec{x})\right),
\end{equation}         
with
$G(\vec{y}-\vec{x})$ satisfying $\nabla^2 G(\vec{y}-\vec{x})=\delta^3
(\vec{y}-\vec{x})$. To our knowledge, it does not appear obvious how one 
might 
incorporate the above non-local and non-Lorentz-covariant transformation 
into the Vilkovisky-DeWitt formulation. This is an issue deserving 
further investigations.   
   
\acknowledgments
We thank W.-F. Kao for discussions.
This work is supported in part by
National Science Council of R.O.C. under grant numbers 
NSC 87-2112-M-009-038, and NSC 88-2112-M-009-002.       

\newpage

\appendix
\section{The Higgs-boson Self-energy and Vilkovisky-DeWitt Effective Action }

In this Appendix, we calculate the Higgs-boson self-energy
of scalar QED from the Vilkovisky-DeWitt effective action.
We will focus on the momentum-dependent part of the self-energy,
which is not a part of the effective potential calculated in Sec. II.
Furthermore only the infinite part of the self-energy will be calculated.
We thus perform the calculation in the symmetry phase of the theory. 

We begin with the Lagrangian in Eq. (4) where $\mu^2$ is negative, i.e., 
$-\mu^2\equiv u^2 >0$. In this case, $\phi_1$ and $\phi_2$ have an identical
mass
$m_{\phi}^2=2\lambda u^2$. Let us rename $\phi_1$ as $\rho$ and $\phi_2$
as $\chi$ according to our notation in the symmetry-broken phase. 
If one follows the background field expansion
in Eq. (2), one would expand the QED action by writing 
$A^{\mu}=A^{\mu}_{cl}+B^{\mu}$, $\rho=\rho_{cl}+
\eta_1$, and $\chi=\chi_{cl}+\eta_2$,
with $A^{\mu}_{cl}$, $\rho_{cl}$ and $\chi_{cl}$ the 
classical background fields, and
$B^{\mu}$, $\eta_1$ and $\eta_2$ the 
corresponding quantum fluctuations. 
However, as mentioned earlier, the above quantum fluctuations 
should be replaced by vectors $\sigma^i$ in the configuration space.
Hence the action in Eq. (2) should be expanded covariantly
\cite{vil} in powers of
$\sigma^i$. To simplify our notations,
we use $\tilde{B}_{\mu}$ and $\tilde{\eta}_i$ to denote the new quantum
fluctuations.  
Since we  will only calculate the self-energy of $\rho$, 
we may take $A^{\mu}_{cl}=\chi_{cl}=0$ for simplicity.
With the covariant expansion, the Lagrangian in Eq. (4) generate 
the following quadratic terms:
\begin{eqnarray}
L_{quad}=&-&{1\over 4}(\partial_{\mu}\tilde{B}_{\nu}-\partial_{\nu}
\tilde{B}_{\mu})^2 
+{1\over 2}(\partial_{\mu}\tilde{\eta}_1)(\partial^{\mu}\tilde{\eta}_1) 
+{1\over 2}(\partial_{\mu}\tilde{\eta}_2)
(\partial^{\mu}\tilde{\eta}_2)\nonumber \\
&+&e\rho_{cl}(\partial^{\mu}\tilde{\eta}_2)\tilde{B}_{\mu}
-e\tilde{\eta}_2(\partial^{\mu}\rho_{cl})
\tilde{B}_{\mu}+{1\over 2}e^2\rho_{cl}^2\tilde{B}^{\mu}\tilde{B}_{\mu}
\nonumber \\
&-&\lambda\left[{1\over 2}\rho_{cl}^2(3\tilde{\eta}_1^2+
\tilde{\eta}_2^2)+
u^2(\tilde{\eta}_1^2+\tilde{\eta}_2^2)\right]\nonumber \\
&-&\Gamma^{\Phi^l}_{\Phi^m \Phi^n}{\delta S\over \delta \Phi^l}
\tilde{\phi}^m\tilde{\phi}^n,
\end{eqnarray}   
where $\Phi^l$ and $\tilde{\phi}^l$ denote generically the classical 
background fields and the quantum fluctuations respectively.  
We choose the 
$R_{\xi}$ background-field gauge with the gauge-fixing term:
\begin{equation}
L_{gf}=-{1\over 2\alpha}(\partial_{\mu}\tilde{B}^{\mu}-
\alpha e\rho_{cl}\tilde{\eta}_2)^2.
\end{equation}
The corresponding Faddeev-Popov Lagrangian is then
\begin{equation}
L_{FP}=\omega^*(-\partial^2-\alpha e^2\rho_{cl}^2)\omega.
\end{equation}
Compared to the usual background-field formalism, the quadratic quantum
flucuations $L_{quad}$ contain extra terms proportional to the connection
$\Gamma^i_{jk}$ of the configuration space.
These extra terms are crucial for the cancellation of gauge-parameter
dependence in the Higgs-boson self-energy.
From Eqs. (5), (6), (7) and (8), we calculate those connections which
are relevant to the Higgs-boson self-energy. We find
\begin{eqnarray}
&&\Gamma^{\rho(z)}_{A_{\mu}(x)A_{\nu}(y)}\vert_{\Phi}
=-e^2\rho_{cl}(z)(\partial^{\mu}_x N^{xz})(\partial^{\nu}_y N^{yz})\nonumber
\\
&&\Gamma^{\rho(z)}_{\chi(x)\chi(y)}\vert_{\Phi}= e^2N^{xy}(\delta^4(y-z)+
\delta^4(x-z))
\rho_{cl}(z)-e^4\rho_{cl}(z)N^{zx}N^{zy}\rho_{cl}(x)\rho_{cl}(y)\nonumber \\
&&\Gamma^{\rho(z)}_{A_{\mu}(x)\chi(y)}\vert_{\Phi}=e(\partial^{\mu}_x N^{zx})
\delta^4(z-y)-e^3\rho_{cl}(z)(\partial^{\mu}_xN^{xz})N^{zy}\rho_{cl}(y),
\end{eqnarray}
where $N^{xy}=<x\vert {1\over \partial^2+e^2\rho_{cl}^2(X)}\vert y>$
with $X_{\mu}\vert x>=x_{\mu}\vert x>$; and 
the notation $\vert_{\Phi}$ denotes evaluating
the connection at the classical background fields. 
The above connections are to be
multiplied by ${\delta S\over \delta \rho}\vert_{\Phi}\equiv 
(-\partial^2-2\lambda u^2-\lambda \rho_{cl}^2)\rho_{cl}$ with the 
space-time variable $z$ integrated over. It is interesting to note 
that the product of $\Gamma$ and ${\delta S\over \delta \rho}$ 
contain terms which are able to generate the Higgs-boson self-energy. 
For example, in the expression $-\int d^4x d^4y 
(\Gamma^{\rho(z)}_{A_{\mu}(x)A_{\nu}(y)}{\delta S\over \delta \rho}
\vert_{\Phi}) 
\tilde{B}_{\mu}(x)\tilde{B}_{\nu}(y)$, we can set $\rho_{cl}=0$ in $N^{xz}$
and $N^{yz}$ and contract the pair of gauge fields. This gives rise to,
in the momentum space, the following Higgs-boson
self-energy
\begin{equation}
\Sigma_{\rho}^{AA}(p^2)=-{\alpha e^2\over 8\pi^2}{1\over \epsilon'}
p^2.
\end{equation}
where $1/\epsilon'\equiv 1/\epsilon+{1\over 2}\gamma_E-{1\over 2}\ln(4\pi)$
with $\epsilon=n-4$.
For $-\int d^4x d^4y 
(\Gamma^{\rho(z)}_{\chi(x)\chi(y)}{\delta S\over \delta \rho}
\vert_{\Phi}) 
\tilde{\eta}_2(x)\tilde{\eta}_2(y)$, we again set $\rho_{cl}=0$ in $N^{xy}$
and contract the pair of scalar fields. We obtain the Higgs-boson self-energy
\begin{equation}
\Sigma_{\rho}^{\chi\chi}(p^2)={e^2\over 4\pi^2}{1\over \epsilon'}
p^2.
\end{equation}   
Finally, the term $-\int d^4x d^4y 
(\Gamma^{\rho(z)}_{A_{\mu}(x)\chi(y)}{\delta S\over \delta \rho}
\vert_{\Phi}) 
\tilde{B}_{\mu}(x)\tilde{\eta}_2(y)$ can produce an effective 
$\rho_{cl}-\tilde{B}_{\mu}-\tilde{\eta}_2$ vertex, namely,
$\int {d^4p d^4k\over (2\pi)^8}\Gamma^{\mu}(p,k)\rho_{cl}(k)
\tilde{B}_{\mu}(p)\eta_2(-p-k)$, with 
$\Gamma_{\mu}(p,k)=i{p^{\mu}\over p^2}\cdot 
(k^2-2\lambda u^2)$.  
This vertex can contribute to the Higgs-boson self-energy by contracting with
another vertex of the same kind. Similarly, it could contract
with an ordinary $\rho_{cl}-\tilde{B}_{\mu}-\tilde{\eta}_2$ vertex. 
It turns out that 
both contractions produce only finite contributions to the 
momentum-dependent part of the Higgs-boson self-energy.  
Therefore Eqs. (A5) and (A6) are the only divergent contributions 
of $\Gamma^i_{jk}$ to the momentum-dependent part of the Higgs-boson
self-energy. These contributions are to be added to the self-energy obtained
by contracting a pair of 
ordinary $\rho_{cl}-\tilde{B}_{\mu}-\tilde{\eta}_2$ vertices.
We find
\begin{equation}
\Sigma_{\rho}^{ordinary}(p^2)=
{e^2\over 8\pi^2}{1\over \epsilon'}(3+\alpha)
p^2.
\end{equation} 
From Eqs. (A5), (A6) and (A7), we arrive at
\begin{equation}
\Sigma_{\rho}(p^2)=\Sigma_{\rho}^{AA}(p^2)
+\Sigma_{\rho}^{\chi\chi}(p^2)
+\Sigma_{\rho}^{ordinary}(p^2)={5e^2\over 8\pi^2}{1\over \epsilon'}
p^2.
\end{equation}
We can see that the gauge-parameter dependence of 
$\Sigma_{\rho}^{ordinary}(p^2)$ is cancelled by 
that of $\Sigma_{\rho}^{AA}(p^2)$! From Eq. (A8),
the wave-function renormalization
constant of the Higgs boson is found to be
\begin{equation}
Z_{\rho}=1-{5e^2\over 8\pi^2}{1\over \epsilon'}.
\end{equation}
This result can be applied to the model in Sec. III with the replacement 
$e\to -{g_1\over 2}$
according to Eq. (17). Hence $Z_{\rho}=1-
{5g_1^2\over 32\pi^2}{1\over \epsilon'}$ in that model, which reproduces 
the relevant part of
Eq. (\ref{renc}) calculated in the Landau-DeWitt gauge. 


\begin{references}
\bibitem{com1}
We should point out that
the simple replacement of $\eta^i$ with $\sigma^i(\Phi,\phi)$, as suggested
in Ref.\cite{vil}, is not satisfactory for calculations beyond the one-loop
order. 
An effective action obtained by such a prescription 
does not generate one-particle irreducible
diagrams at 
the higher loops. A modified construction was proposed by DeWitt 
in Ref.\cite{dw2}. However, since we are only concerning with one-loop 
corrections, we shall adhere to the current construction.
\bibitem{com2}
Although properties of this gauge are discussed in Ref.\cite{FT}, 
the authors of this paper used another 
gauge to calculate the effective potential of scalar QED.
\bibitem{grassmann}
In the current gauge, $\tilde{D}^{-1}_{ij}=D^{-1}_{ij}$ 
since $\Gamma^i_{jk}$ vanishes. Furthermore,
for ghost fields, it is the ghost propagator, rather than its inverse, 
that will appear in
the effective action. This is due to the Grassmannian nature 
of ghost fields.
\bibitem{fermion}
We have suppressed the fermionic part of the gauge-fixing term. The
justification of this simplified treatment is discussed in
the first paper of Ref.\cite{rebhan}.
\end{references}
\end{document}